\documentclass[iop,apjl]{emulateapj}

\usepackage{graphicx}

\usepackage{natbib}
\bibpunct{(}{)}{;}{a}{}{,}
\newcommand{\msunyr}{\ensuremath{\mathit{M}_{\odot}{\rm yr}^{-1}}}   
\newcommand{\kms}{\ensuremath{{\rm km\,s^{-1}}}}                   
\newcommand{\K}{\mathrm{K}}
\newcommand{\lsun}{\ensuremath{\mathit{L}_{\odot}}}                  

\newcommand{\lstar}{\ensuremath{\mathit{L}_{\star}}}                 
\newcommand{\mdot}{\ensuremath{\dot{M}}}                             
\newcommand{\teff}{\ensuremath{\mathit{T}_{\rm eff}}}                
\newcommand{\vinf}{\ensuremath{v_{\infty}}}                          

\newcommand{\phiorb}{\ensuremath{\phi_{\mathrm{orb}}}}                 
\newcommand{\etaa}{\ensuremath{\eta_{\mathrm{A}}}}                 
\newcommand{\etab}{\ensuremath{\eta_{\mathrm{B}}}}                 

\newcommand{\hststis}{{\it HST}/STIS}                          

\shorttitle{Latitude-dependent effects in Eta Car caused by the companion}
\shortauthors{Groh et al.}

\begin{document}

\title{A companion as the cause of latitude-dependent effects in the wind of Eta Carinae \altaffilmark{1}}
\author{J. H. Groh\altaffilmark{2,3}, T. I. Madura\altaffilmark{3}, D. J. Hillier\altaffilmark{4},  C. J. H. Kruip\altaffilmark{5}, and G. Weigelt\altaffilmark{3}}
\email{jose.groh@unige.ch}

\altaffiltext{1}{Based on observations made with HST/STIS.}
\altaffiltext{2}{Geneva Observatory, Geneva University, Chemin des Maillettes 51, CH-1290 Sauverny, Switzerland}
\altaffiltext{3}{Max-Planck-Institut f\"ur Radioastronomie, Auf dem H\"ugel 69, D-53121 Bonn, Germany}
\altaffiltext{4}{Department of Physics and Astronomy, University of Pittsburgh, 3941 O'Hara Street, Pittsburgh, PA, 15260, USA}
\altaffiltext{5}{Leiden Observatory, Leiden University, Postbus 9513, 2300 RA Leiden, The Netherlands}
\begin{abstract}

We analyze spatially resolved spectroscopic observations of the Eta Carinae binary system obtained with HST/STIS. Eta Car is enshrouded by the dusty Homunculus nebula, which scatters light emitted by the central binary and provides a unique opportunity to study a massive binary system from different vantage points. We investigate the latitudinal and azimuthal dependence of H$\alpha$ line profiles caused by the presence of a wind-wind collision (WWC) cavity created by the companion star. Using two-dimensional radiative transfer models, we find that the wind cavity can qualitatively explain the observed line profiles around apastron. Regions of the Homunculus which scatter light that propagated through the WWC cavity show weaker or no H$\alpha$ absorption. Regions scattering light that propagated through a significant portion of the primary wind show stronger P Cygni absorption. Our models overestimate the  H$\alpha$ absorption formed in the primary wind, which we attribute to photoionization by the companion, not presently included in the models. We can qualitatively explain the latitudinal changes that occur during periastron, shedding light on the nature of Eta Car's spectroscopic events. Our models support the idea that during the brief period of time around periastron when the primary wind flows unimpeded toward the observer, H$\alpha$ absorption occurs in directions toward the central object and Homunculus SE pole, but not toward equatorial regions close to the Weigelt blobs. We suggest that observed latitudinal and azimuthal variations are dominated by the companion star via the WWC cavity, rather than by rapid rotation of the primary star.

\end{abstract}


\section{Introduction}  \label{intro}

Eta Carinae is a rarity among the rarities that constitute massive stars in the short, unstable Luminous Blue Variable (LBV) phase. The nature and fate of the central object are still the subject of intense debates, although consensus seems to exist around an eccentric massive binary scenario, as initially proposed by \citet{dcl97}. This scenario is supported by multi-wavelength observations from X-rays \citep{corcoran10} to radio wavelengths \citep{duncan03}.

The ultraviolet and optical spectra are strongly dominated by the dense wind of the primary star (hereafter \etaa), hinting that most of the luminosity of the system ($\lstar\simeq5\times10^6~\lsun$, \citealt{dh97}) comes from \etaa\ \citep[][hereafter G12]{hillier01,hillier06, ghm12}. Recent spectroscopic analysis suggests that \etaa\ has a mass-loss rate of $8.5\times10^{-4} \msunyr$ and wind terminal velocity of $\sim420~\kms$ (G12; see also \citealt{hillier01,hillier06}). The companion star (hereafter \etab) has yet to be observed directly, with only indirect constraints available on its temperature and luminosity ($\teff\simeq36,000-41,000~\K$ and $10^5 \lsun \la \lstar \la 10^6 \lsun$; \citealt{mehner10}), and wind properties ($\vinf\sim3000~\kms$ and $\mdot \sim1.4 \times 10^{-5}~\msunyr$; e.g., \citealt{parkin11}).

\defcitealias{ghm12}{G12}
\defcitealias{smith03}{S03}

\begin{figure}
\resizebox{\hsize}{!}{\includegraphics{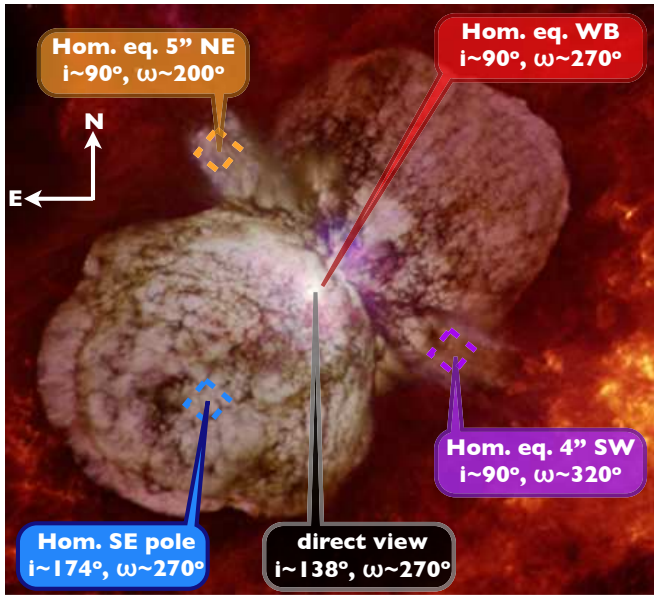}}
\caption{\label{offsetpos}{Offset positions discussed in this Letter. The angles $i$ and $\omega$ correspond approximately to the equivalent orbital inclination (i.\,e. latitude) and longitude of periastron (i.\,e. azimuth) that light scattered off different regions of the Homunculus see the central binary system. The background image shows a $\sim20\arcsec$ box around the central source in Eta Car (credit: Nathan Smith (U. Arizona), NASA).}}
\end{figure}

Eta Car has interested astronomers ever since its dramatic brightness increase in the 1830s. This eruptive event caused the ejection of tens of solar masses of material \citep{smith03b}, enshrouding the central object within a dusty nebula (the Homunculus). The expanding dust residing in the Homunculus scatters light emitted by the central source. The 3D determination of the Homunculus' shape \citep{davidson01,smith06} allows one to relate the scattered light at a given position in the Homunculus to a line-of-sight that views the stellar system from a certain latitude \citep[][hereafter S03]{smith03} and azimuth. In the context of a binary, this implies that different positions in the Homunculus see the binary as if viewed from different orbital inclinations $i$ and longitudes of periastron $\omega$ (Fig. \ref{offsetpos}).

Spatially resolved observations of Eta Car show that, during most phases of the orbital cycle, H and \ion{Fe}{2} spectral lines are latitude dependent. The P Cygni absorption is stronger in spectra taken around the pole of the Homunculus SE lobe than in the spectra of the central object \citep[][K. Nielsen 2009, priv. comm.]{hillier92,smith03, weis05,stahl05,mehner12}. The P Cygni absorption seems to extend to higher velocities in spectra reflected off the Homunculus SE pole, which, combined with the stronger absorption, has been interpreted as evidence for a denser, faster polar wind generated by the rapid rotation of \etaa\ \citepalias{smith03}. The latitudinal dependence  of H and \ion{Fe}{2} seems to be less pronounced around periastron, suggesting that \etaa's wind becomes more spherical \citepalias{smith03}.

However, H and \ion{Fe}{2} lines, which are formed in the wind of \etaa, are extremely affected by the presence of a cavity created by \etab\ \citepalias{ghm12}. This cavity arises naturally in hydrodynamical models as a consequence of the carving of the wind of \etaa\ by the fast, thin wind of \etab\ \citep{pc02}.  Taking into account the presence of  the low-density cavity, two-dimensional (2D) radiative transfer models of Eta Car show that H and \ion{Fe}{2} lines become latitude- and azimuth-dependent \citepalias{ghm12}. This occurs because of the varying amount of primary wind material towards different latitudes and azimuths. These models reproduce the absence of P Cygni absorption in H$\alpha$ and \ion{Fe}{2} lines in spectra seen in the direct view to the central object (which is viewed through the rarified cavity), {\it without} evoking rapid rotation of \etaa.

Can we still consider \etaa\ as a rapid rotator and prototype of massive stars with dense polar winds? The goal of this Letter is to investigate the origin of the observed latitudinal and azimuthal dependencies of \etaa's wind and, in particular, whether they can be explained by the presence of a WWC cavity in the wind of \etaa.

\section{Observations and modeling} \label{obs}

To illustrate our findings, we focus on archival \hststis\ observations obtained around apastron (2000 Mar 20, orbital phase $\phi=10.409$\footnote{We assume the ephemeris from \citet{damineli08_multi}, the orbital cycle labeling from \citet{gd04}, and that periastron occurs at $\phi=0$.}, $52\arcsec \times 0\farcs2$ aperture) and periastron (1998 Mar 20, $\phi=10.046$, $52\arcsec \times 0\farcs1$ aperture). Both datasets were acquired with the G750M grating, covering the region around H$\alpha$ with $R\sim8000$. The spectra were extracted using custom IDL routines using an aperture of 6 half-pixels (0\farcs152) on the central object and close to the Weigelt blobs (WBs; \citealt{weigelt86}, located 0\farcs15--0\farcs30 NW of the central source), and 12 half-pixels elsewhere.  The spectra were corrected for the velocity shift introduced by the dust scattering using narrow forbidden lines as proxies, following \citet{mehner12}. This technique could not be applied to the offsets at 5\arcsec NE and 4\arcsec SW, and the profiles were aligned on the {\it red} side of H$\alpha$, where the influence from \etab\ is minimized \citepalias{ghm12}. Note that this does not affect our conclusions. We refer the reader to e.\,g. \citetalias{smith03} for details about the observations and data reduction.

The aforementioned observations are analyzed using the 2D radiative transfer models of Eta Car from \citetalias{ghm12}. The 2D models take into account \etaa\ and the presence of a rarified WWC cavity and a dense wind-wind interacting region, corresponding to the post-shocked primary wind. We refer the reader to \citetalias{ghm12} for an extensive discussion about the impact of the wind-wind collision (WWC) cavity on the spectrum, and to their table 2 for the full model parameters. We assume an orbital orientation with $i=138\degr$, $\omega=270\degr$, $\mathrm{PA}_z=312\degr$ \citep{madura11}, and half-opening angle of the cavity of $\alpha=57\degr$ \citepalias{ghm12}. This corresponds to the direct view of the central source. Since offset positions in the Homunculus see the binary system under different vantage points, 2D models with different $i$ and $\omega$ were computed for each offset analyzed here.

For visualization purposes, we also present 3D renderings of Smoothed Particle Hydrodynamics (SPH) simulations of the Eta Car binary system, similar to those presented in \citet{mg12}.

\begin{figure}
\resizebox{1.0\hsize}{!}{\includegraphics{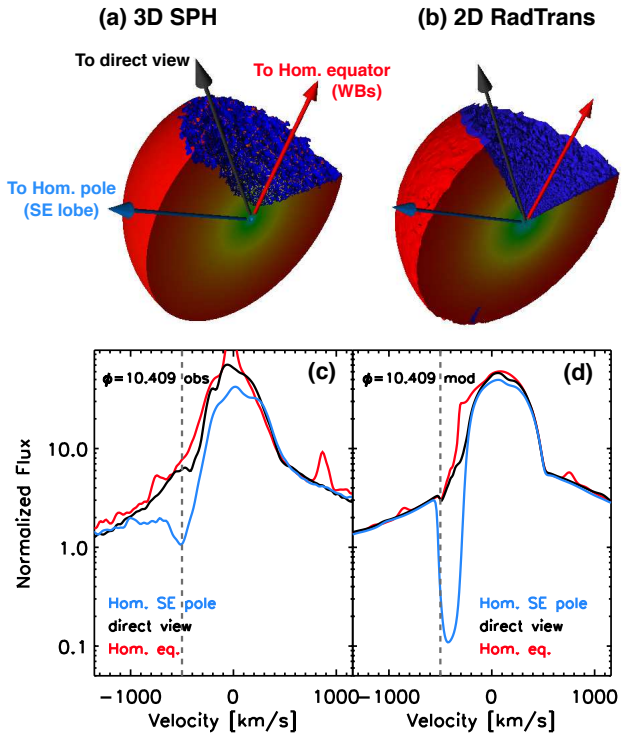}}
\caption{\label{latiap}{{\it a}: 3D rendering of hydrodynamical simulations showing the extended, optically-thick primary wind (cyan/green/yellow/red) and the WWC zone (dark blue) at orbital phase $\phi=10.409$. Material in the primary wind is color coded to radius, i.e., red corresponds to material at larger radii ($\sim75-150$ AU) from the central stars, while green indicates material near the apex of the WWC region ($< 15$ AU from \etaa\ for these phases). Material below the orbital plane has been removed, and the 3D rendering slightly tilted, to allow better visualization of the dynamics of the inner regions. The physical scale corresponds to $\sim \pm 150 AU$ centered on the center of mass of the binary system. {\it b}: Similar to {\it (a)}, but for the 2D radiative transfer models. {\it c}: Observed spectra obtained at $\phi=10.409$ at the Homunculus equator (red), direct view at 138\degr (black), and Homunculus SE pole (blue). To aid the comparison between model and observations, the vertical gray dashed line corresponds to $v=-500~\kms$. {\it d}: Model spectra corresponding to these viewing angles. Note that the narrow emission seen in {\it c} around $-40~\kms$ is due to nebular emission from the Weigelt blobs, and not included in the modeling.}}
\end{figure}

\section{Latitudinal changes in line profiles due to the WWC cavity} \label{apastron}
We first investigate the strong latitudinal changes in line profiles around apastron. In the WWC cavity scenario, the latitudinal variations of line profiles  depend on the amount of primary wind material towards a certain latitude \citepalias{ghm12}. This is regulated mainly by the size of the line formation region compared to the distance of the WWC apex to \etaa, meaning that certain lines (H$\alpha$, \ion{Fe}{2}) are more affected by the cavity than others (e.g., higher Balmer lines). For brevity, we present here results for H$\alpha$, which is formed at a  distance of $\sim10-100$ AU from \etaa. Similar conclusions would be obtained for H$\beta$ and \ion{Fe}{2} lines.

Figure \ref{latiap}c,d presents the observed and 2D model spectra of Eta Car for different viewing angles, corresponding to a direct view of the stellar system (black line), a view from the Homunculus SE pole (light blue), and from the Homunculus equator towards the WBs (red). Our 2D models qualitatively reproduce the observed variation of  H$\alpha$ with latitude, suggesting that the latitudinal changes can be explained by the presence of the WWC cavity. For viewing angles corresponding to line-of-sights that view \etaa\ through the rarified cavity, there is little or no P Cygni absorption because of the reduced density within the cavity. This drastically reduces the $n=2$ level population of H, reducing the H$\alpha$ absorption. This is the case for the direct view to the central source or to the Homunculus equator towards the WBs (Fig. \ref{latiap}a). There is also a fair quantitative agreement between observations and 2D models at these two positions.

\begin{figure}
\resizebox{\hsize}{!}{\includegraphics{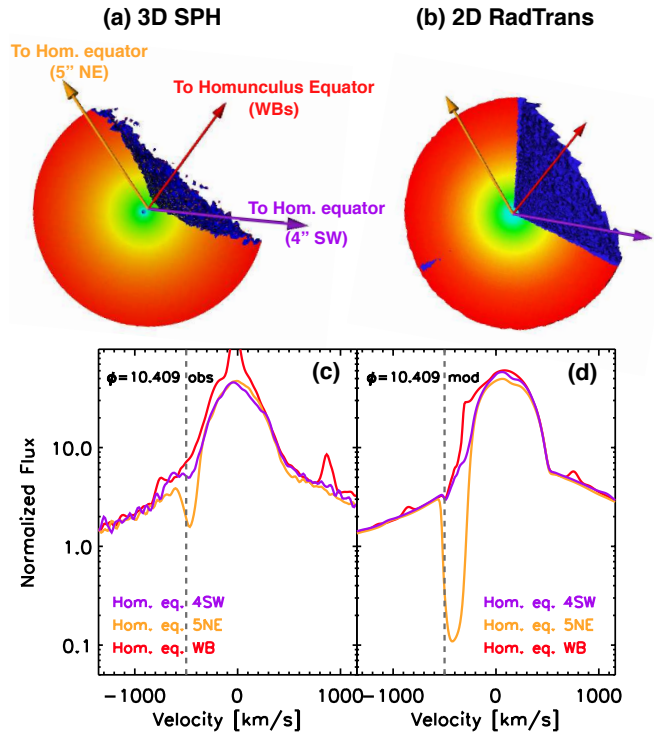}}
\caption{\label{aziap}{Similar to Fig. \ref{latiap}, but for azimuthal variations.}}
\end{figure}

For line-of-sights that cross the undisturbed wind of \etaa, such as for the spectra reflected on the Homunculus SE pole, the population of the $n=2$ energy level of H is unchanged in our model compared to a spherically symmetric case. This results in stronger absorption of the stellar continuum in spectra scattered off the Homunculus SE pole than in spectra of the direct view to the central source, in agreement with the observations. However, our 2D model  overestimates the amount of P Cygni absorption seen in spectra scattered off  the Homunculus SE pole, and underestimates the maximum velocity of P Cygni absorption in H$\alpha$ by $\sim70~\kms$ (blue lines in Fig. \ref{latiap}c,d).

Our favored explanation for these discrepancies is that the polar regions of \etaa's wind could be mildly photoionized by \etab\ \citep{madura11, kruip11}. This effect is not included in our 2D models and would reduce the population of the $n=2$ energy level of H and thus the amount of H$\alpha$ absorption. In addition, photoionization by \etab\ would likely change the driving of the wind of \etaa, perhaps allowing it to reach velocities slightly higher than regions of \etaa's wind not affected by \etab.

Alternatively, a wrong choice of stellar and wind parameters could also be the culprit, e.\,g. a {\it lower} \mdot\ over the poles would be needed to produce less absorption in H$\alpha$, but the fit to the UV and optical emission lines would be significantly worse \citepalias{ghm12}. In addition, we cannot rule out that the observations do not correspond to the pure polar spectrum, with the intrinsic P Cygni absorption being diluted by continuum and H$\alpha$ emission from the ejecta in the inner arcsecond.

Lastly, we cannot preclude that the wind is slightly faster at the pole because of rapid rotation \citepalias{smith03}, although our 2D cavity models suggest that $\vinf$ over the pole would be much smaller ($\simeq 500~\kms$) than previous observational estimates.

\section{Azimuthal variations due to the WWC cavity} \label{aziapastron}
We turn our attention to azimuthal variations observed in scattered light off the Homunculus around apastron. These would not be expected in a single rapid rotator scenario\footnote{A strong misalignment of the rotation and Homunculus polar axes could cause azimuthal variations  in scattered light  off the Homunculus. However, it would be challenging to produce P Cygni absorption {\it both} at the Homunculus pole and equator, as observed.}, while a WWC cavity intrinsically produces azimuthal density variations (Fig. \ref{aziap}a). Therefore, this crucial difference could allow one to distinguish between the two scenarios.

Figure \ref{aziap}c presents the H$\alpha$ line profiles scattered off the Homunculus equator, around the WBs (red), 5\arcsec NE (orange), and 4\arcsec SW (purple). These positions probe different azimuths, as if the system were viewed from $i=90\degr$, but different $\omega$. The corresponding synthetic H$\alpha$ line profiles from our 2D models are shown in Fig. \ref{aziap}d. Our 2D model qualitatively explains the variations in strength of the P Cygni absorption as a function of azimuth for an orbital orientation with $\omega\sim270\degr$,  $i\sim138\degr$, and $\alpha=57\degr$. Lower values of $\omega$ would be allowed for larger $\alpha$.

The changes in the P Cygni absorption occur because of the variation of primary wind material at different azimuths in the equatorial regions.  Line-of-sights that view the system down the WWC cavity have little primary wind material between the observer and \etaa. This causes less absorption in H$\alpha$. This is the case for the spectra scattered off the WBs (red) and at 4\arcsec SW (purple). Conversely, the spectrum scattered off the Homunculus equator at 5\arcsec NE probes essentially high densities corresponding to the unmodified primary wind. The quantitative agreement of our 2D models is satisfactory at WB and 4\arcsec SW, but overestimates the amount of absorption seen at 5\arcsec NE. This discrepancy resembles that seen in the Homunculus polar spectrum and could also be caused by neglecting photoionization from \etab\ in our 2D models (Sect. \ref{apastron}).

\section{Variations around periastron and the nature of the spectroscopic events} \label{periastron}

Because Eta Car is a highly eccentric binary system, the carving of the primary wind by \etab\ changes as a function of $\phi$. Significant variations are expected around periastron, when the cavity gets closer to \etaa\ and becomes distorted due to the high orbital velocities of \etab\ \citep[e.\,g.][]{okazaki08, parkin11}.

\begin{figure}
\resizebox{\hsize}{!}{\includegraphics{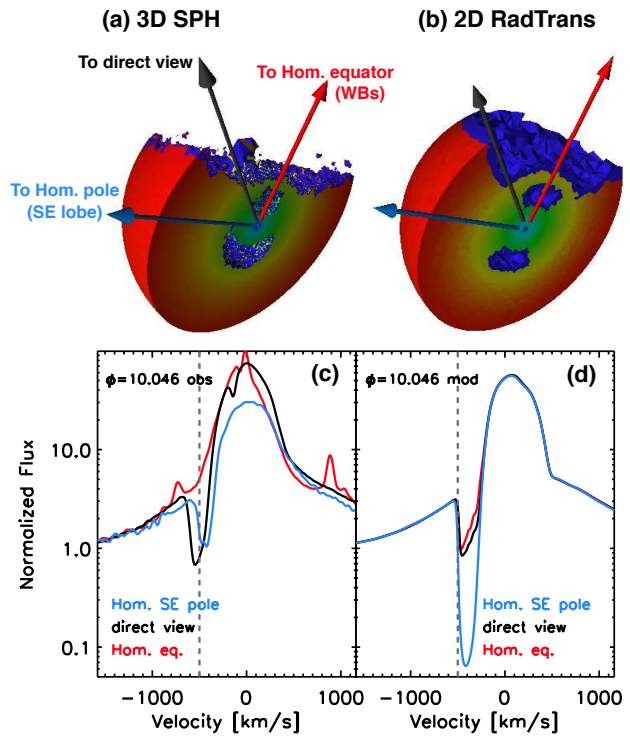}}
\caption{\label{latiperi}{Similar to Fig. \ref{latiap}, but for just after periastron (\phiorb=10.046).}}
\end{figure}

Figure \ref{latiperi}a shows a 3D rendering of the primary wind and WWC cavity at $\phi=10.046$. Its morphology and orientation with respect to the observer are greatly modified compared to apastron (Fig. \ref{latiap}a).  The observed line profiles are still latitude-dependent, with the key difference being the appearance of P Cygni absorption in the spectrum of the direct view to the central object (Fig. \ref{latiperi}c). Notice that the spectrum scattered off the WBs (red) does not show H$\alpha$ absorption near periastron \citep{gull09}.

The hydrodynamics and morphology of the WWC cavity around periastron are intrinsically 3D and, as such, our 2D modeling fails to fully reproduce the observed changes at all latitudes and azimuths. However, as important insights could be obtained with a 2D model, we present here a first attempt to model the periastron spectra at different latitudes.

Figure \ref{latiperi}d displays the synthetic H$\alpha$ line profiles from our 2D model, assuming the density structure depicted in Fig.  \ref{latiperi}b. After periastron, for a short amount of time the wind of \etaa\ is able to escape to regions previously occupied by the WWC cavity, up to a distance of $\sim60$~AU at \phiorb=10.046. This material increases the column density of neutral H, in particular for viewing angles with $i$ corresponding to the direct view. Our 2D models indicate that this increase in primary wind material is enough to cause H$\alpha$ P Cygni absorption (Fig. \ref{latiperi}d).

This portion of primary wind material, which causes an increase in column density, could be easily confused with the {\it shell ejection} scenario advocated by earlier studies \citep[][S03]{davidson99}. However, the key difference is that no change in $\mdot$, nor instability in \etaa, are needed to eject the required portion of primary wind material in line-of-sight to \etaa. It is simply the result of the dynamical interplay between the primary and secondary wind in this eccentric system \citep{madura11}. For lines formed in \etaa, we find that the portion of primary wind that flows unimpeded for a brief period of time after periastron has a significant impact on the line profiles.

The fits to spectra scattered at the Homunculus SE pole suffer from a problem similar to that seen at apastron, when the P Cygni absorption is overestimated by our 2D model. However, our model can qualitatively reproduce the velocities of the P Cygni absorption seen in polar and direct view spectra. In addition, the 2D models predict little variability in the polar spectrum between apastron and periastron, in agreement with the observations \citep{stahl05,mehner12}. The P Cygni absorption at the equator is also overestimated by the 2D model. We attribute this to the model assumption breaking down at this latitude, since the density structure of the primary wind is severely modified by the WWC cavity (Fig. \ref{latiperi}). Namely, the 3D SPH simulations predict three structures in line-of-sight to \etaa\ when seen from the equator: a fossil outer cavity, a geometrically thin ($\sim20$ AU) portion of primary wind, and a newly-formed cavity by \etab. The H$\alpha$ absorption arising in our model  comes from the thin portion of primary wind material, but the assumed source function and opacities are unrealistic. This is  because the newly formed cavity allows photons from \etaa\ and \etab\ to ionize this thin portion of primary wind in the equatorial regions. Because of our extremely simplified assumptions here, it is mandatory to include 3D and ionization effects to improve the modeling at periastron.

\section{Implications: a companion that does it all?}

We showed that the 2D models of Eta Carinae from \citetalias{ghm12} qualitatively explain the latitudinal and azimuthal variations of H$\alpha$ line profiles observed around apastron. We also attempted a first modeling of the variations seen during periastron, showing promising results and explaining the increase in H$\alpha$ P Cygni absorption observed in the direct view to the central object after periastron.

Latitudinal variations in P Cygni absorption lines have been interpreted as being caused by the rapid rotation of \etaa, which would produce a dense polar wind \citepalias{smith03}. We suggest that an otherwise spherical \etaa\ wind carved by \etab\ can explain these variations. The fact that the presence of \etab\ affects the H$\alpha$ and \ion{Fe}{2} line profiles \citepalias{ghm12}, which have been used as diagnostics of rapid rotation, hampers the determination of how rapid \etaa\ is spinning. Ultimately, this also raises the point of whether rapid rotation is really present.

More importantly, our 2D models reproduce not only the latitudinal, but also the azimuthal variations observed in spectra scattered off the Homunculus equator. Azimuthal changes are in principle not expected from  rapid rotation, and additional mechanisms would have to be evoked to explain the observations. Our cavity scenario naturally produces azimuthal variations, caused by different amounts of primary wind material at different azimuths, which is controlled by the opening angle of the WWC cavity. {\it The results from this Letter cast serious doubts on the idea that \etaa\ has been a rapid rotator in recent decades.}

Our 2D models can qualitatively explain the latitudinal dependence variations seen in the ground-based H$\alpha$ line profiles of \citet{mehner12} (their Fig. 8). Our 2D models also suggest that the spectrum reflected off the Homunculus pole should remain essentially unchanged during periastron, in excellent agreement with the observations. The lack of significant long-term variation in the polar \citep{mehner12} and equatorial spectrum reflected off the WBs \citep{gull09}, both in line strengths and continuum flux, also would argue against dramatic long-term changes in stellar parameters of \etaa. 

Finally, one might wonder how Eta Car fits in the context of stellar evolution through the LBV phase. It might be that AG Car or HR Car \citep{gdh09,ghd09}, which so far do not have detected companions, might be more appropriate prototypes for Galactic LBVs rather than Eta Car. Also, how does Eta Car fare in the context of the most massive stars, such as those observed in the core of the massive stellar clusters NGC 3603 and R136 \citep{crowther10}? Is Eta Car an evolved version of those stars, at a stage when a Giant Eruption recently occurred? Or is binary evolution required to reach the stage where Eta Car is today? Specially if the presence of \etab\ turns out to be a necessary factor for the occurrence of the Giant Eruption,  perhaps the main observational properties that make Eta Car so unique in the Galaxy are directly or indirectly linked to the presence of a binary companion.

Our results provide the basic foundation to understand line profile variations scattered off the Homunculus, and to disentangle the long-term evolution of the system from phase-locked changes. The presence of a WWC cavity seems to be the dominant effect to understand H$\alpha$ on the central source and WBs. We encourage future theoretical efforts to take into account the ionization from \etab, which is expected to play a role at selected orbital phases and latitudes/azimuths, and to explain the behavior of other lines such as H$\delta$ and \ion{He}{1}. 

\acknowledgments

JHG is supported by an Ambizione Fellowship of the Swiss National Science Foundation. JHG and TIM thank the Max Planck Society for partial financial support. We acknowledge useful discussions with A.~Damineli, T.~Gull, and S.~Owocki.

{\it Facilities:} \facility{HST/STIS}


\begin{thebibliography}{34}
\expandafter\ifx\csname natexlab\endcsname\relax\def\natexlab#1{#1}\fi

\bibitem[{{Corcoran} {et~al.}(2010){Corcoran}, {Hamaguchi}, {Pittard},
  {Russell}, {Owocki}, {Parkin}, \& {Okazaki}}]{corcoran10}
{Corcoran}, M.~F., {Hamaguchi}, K., {Pittard}, J.~M., {et~al.} 2010, \apj, 725,
  1528

\bibitem[{{Crowther} {et~al.}(2010){Crowther}, {Schnurr}, {Hirschi}, {Yusof},
  {Parker}, {Goodwin}, \& {Kassim}}]{crowther10}
{Crowther}, P.~A., {Schnurr}, O., {Hirschi}, R., {et~al.} 2010, \mnras, 408,
  731

\bibitem[{{Damineli} {et~al.}(1997){Damineli}, {Conti}, \& {Lopes}}]{dcl97}
{Damineli}, A., {Conti}, P.~S., \& {Lopes}, D.~F. 1997, New Astronomy, 2, 107

\bibitem[{{Damineli} {et~al.}(2008){Damineli}, {Hillier}, {Corcoran}, {Stahl},
  {Groh}, {Arias}, {Teodoro}, {Morrell}, {Gamen}, {Gonzalez}, {Leister},
  {Levato}, {Levenhagen}, {Grosso}, {Colombo}, \&
  {Wallerstein}}]{damineli08_multi}
{Damineli}, A., {Hillier}, D.~J., {Corcoran}, M.~F., {et~al.} 2008, \mnras,
  386, 2330

\bibitem[{{Davidson}(1999)}]{davidson99}
{Davidson}, K. 1999, in Astronomical Society of the Pacific Conference Series,
  Vol. 179, Eta Carinae at The Millennium, ed. {J.~A.~Morse, R.~M.~Humphreys,
  \& A.~Damineli}, 304

\bibitem[{{Davidson} \& {Humphreys}(1997)}]{dh97}
{Davidson}, K. \& {Humphreys}, R.~M. 1997, \araa, 35, 1

\bibitem[{{Davidson} {et~al.}(2001){Davidson}, {Smith}, {Gull}, {Ishibashi}, \&
  {Hillier}}]{davidson01}
{Davidson}, K., {Smith}, N., {Gull}, T.~R., {Ishibashi}, K., \& {Hillier},
  D.~J. 2001, \aj, 121, 1569

\bibitem[{{Duncan} \& {White}(2003)}]{duncan03}
{Duncan}, R.~A. \& {White}, S.~M. 2003, \mnras, 338, 425

\bibitem[{{Groh} \& {Damineli}(2004)}]{gd04}
{Groh}, J.~H. \& {Damineli}, A. 2004, Information Bulletin on Variable Stars,
  5492, 1

\bibitem[{{Groh} {et~al.}(2009{\natexlab{a}}){Groh}, {Damineli}, {Hillier},
  {Barb{\'a}}, {Fern{\'a}ndez-Laj{\'u}s}, {Gamen}, {Mois{\'e}s}, {Solivella},
  \& {Teodoro}}]{gdh09}
{Groh}, J.~H., {Damineli}, A., {Hillier}, D.~J., {et~al.} 2009{\natexlab{a}},
  \apjl, 705, L25

\bibitem[{{Groh} {et~al.}(2009{\natexlab{b}}){Groh}, {Hillier}, {Damineli},
  {Whitelock}, {Marang}, \& {Rossi}}]{ghd09}
{Groh}, J.~H., {Hillier}, D.~J., {Damineli}, A., {et~al.} 2009{\natexlab{b}},
  \apj, 698, 1698

\bibitem[{{Groh} {et~al.}(2012){Groh}, {Hillier}, {Madura}, \&
  {Weigelt}}]{ghm12}
{Groh}, J.~H., {Hillier}, D.~J., {Madura}, T.~I., \& {Weigelt}, G. 2012, MNRAS, 423, 1623
 

\bibitem[{{Gull} {et~al.}(2009){Gull}, {Nielsen}, {Corcoran}, {Madura},
  {Owocki}, {Russell}, {Hillier}, {Hamaguchi}, {Kober}, {Weis}, {Stahl}, \&
  {Okazaki}}]{gull09}
{Gull}, T.~R., {Nielsen}, K.~E., {Corcoran}, M.~F., {et~al.} 2009, \mnras, 396,
  1308

\bibitem[{{Hillier} \& {Allen}(1992)}]{hillier92}
{Hillier}, D.~J. \& {Allen}, D.~A. 1992, \aap, 262, 153

\bibitem[{{Hillier} {et~al.}(2001){Hillier}, {Davidson}, {Ishibashi}, \&
  {Gull}}]{hillier01}
{Hillier}, D.~J., {Davidson}, K., {Ishibashi}, K., \& {Gull}, T. 2001, \apj,
  553, 837

\bibitem[{{Hillier} {et~al.}(2006){Hillier}, {Gull}, {Nielsen}, {Sonneborn},
  {Iping}, {Smith}, {Corcoran}, {Damineli}, {Hamann}, {Martin}, \&
  {Weis}}]{hillier06}
{Hillier}, D.~J., {Gull}, T., {Nielsen}, K., {et~al.} 2006, \apj, 642, 1098

\bibitem[{{Kruip}(2011)}]{kruip11}
{Kruip}, C. 2011, Ph.~D.~thesis, University of Leiden, Netherlands

\bibitem[{{Madura} \& {Groh}(2012)}]{mg12}
{Madura}, T.~I. \& {Groh}, J.~H. 2012, \apjl, 746, L18

\bibitem[{{Madura} {et~al.}(2012){Madura}, {Gull}, {Owocki}, {Groh}, {Okazaki},
  \& {Russell}}]{madura11}
{Madura}, T.~I., {Gull}, T.~R., {Owocki}, S.~P., {et~al.} 2012, \mnras, 420,
  2064

\bibitem[{{Mehner} {et~al.}(2010){Mehner}, {Davidson}, {Ferland}, \&
  {Humphreys}}]{mehner10}
{Mehner}, A., {Davidson}, K., {Ferland}, G.~J., \& {Humphreys}, R.~M. 2010,
  \apj, 710, 729

\bibitem[{{Mehner} {et~al.}(2012){Mehner}, {Davidson}, {Humphreys},
  {Ishibashi}, {Martin}, {Ruiz}, \& {Walter}}]{mehner12}
{Mehner}, A., {Davidson}, K., {Humphreys}, R.~M., {et~al.} 2012, \apj, 751, 73

\bibitem[{{Okazaki} {et~al.}(2008){Okazaki}, {Owocki}, {Russell}, \&
  {Corcoran}}]{okazaki08}
{Okazaki}, A.~T., {Owocki}, S.~P., {Russell}, C.~M.~P., \& {Corcoran}, M.~F.
  2008, \mnras, 388, L39

\bibitem[{{Parkin} {et~al.}(2011){Parkin}, {Pittard}, {Corcoran}, \&
  {Hamaguchi}}]{parkin11}
{Parkin}, E.~R., {Pittard}, J.~M., {Corcoran}, M.~F., \& {Hamaguchi}, K. 2011,
  \apj, 726, 105

\bibitem[{{Pittard} \& {Corcoran}(2002)}]{pc02}
{Pittard}, J.~M. \& {Corcoran}, M.~F. 2002, \aap, 383, 636

\bibitem[{{Smith}(2006)}]{smith06}
{Smith}, N. 2006, \apj, 644, 1151

\bibitem[{{Smith} {et~al.}(2003{\natexlab{a}}){Smith}, {Davidson}, {Gull},
  {Ishibashi}, \& {Hillier}}]{smith03}
{Smith}, N., {Davidson}, K., {Gull}, T.~R., {Ishibashi}, K., \& {Hillier},
  D.~J. 2003{\natexlab{a}}, \apj, 586, 432

\bibitem[{{Smith} {et~al.}(2003{\natexlab{b}}){Smith}, {Gehrz}, {Hinz},
  {Hoffmann}, {Hora}, {Mamajek}, \& {Meyer}}]{smith03b}
{Smith}, N., {Gehrz}, R.~D., {Hinz}, P.~M., {et~al.} 2003{\natexlab{b}}, \aj,
  125, 1458

\bibitem[{{Stahl} {et~al.}(2005){Stahl}, {Weis}, {Bomans}, {Davidson}, {Gull},
  \& {Humphreys}}]{stahl05}
{Stahl}, O., {Weis}, K., {Bomans}, D.~J., {et~al.} 2005, \aap, 435, 303

\bibitem[{{Weigelt} \& {Ebersberger}(1986)}]{weigelt86}
{Weigelt}, G. \& {Ebersberger}, J. 1986, \aap, 163, L5

\bibitem[{{Weis} {et~al.}(2005){Weis}, {Stahl}, {Bomans}, {Davidson}, {Gull},
  \& {Humphreys}}]{weis05}
{Weis}, K., {Stahl}, O., {Bomans}, D.~J., {et~al.} 2005, \aj, 129, 1694

\end{thebibliography}
\end{document}